\documentclass[aps,prd,10pt,twocolumn]{revtex4}

\usepackage{epsfig}
\usepackage{bm}
\usepackage{latexsym}
\usepackage{natbib}
\usepackage{url}
\usepackage{dcolumn}
\usepackage{color}
\usepackage{amsfonts,amssymb,amsmath}
\usepackage{graphicx,epsfig}
\usepackage{psfrag}
\usepackage{subfigure}
\usepackage{hyperref}
\hypersetup{colorlinks=true}

\newcommand{\ba}{\begin{array}}
\newcommand{\ea}{\end{array}}
\newcommand{\be}{\begin{equation}}
\newcommand{\ee}{\end{equation}}
\newcommand{\bea}{\begin{eqnarray}}
\newcommand{\eea}{\end{eqnarray}}

\def\beq{\begin{equation}}
\def\eeq{\end{equation}}
\def\bea{\begin{eqnarray}}
\def\eea{\end{eqnarray}}

\begin{document}

\title{Anomalous EDGES 21-cm Signal and Moduli Dominated Era}

\author{Mansi Dhuria$^{a,}$\footnote{mansidhuria@iitram.ac.in}}
\affiliation{
$^a$ Institute of Infrastructure Technology Research and Management, Ahmedabad, India \\
}

\begin{abstract}

The EDGES collaboration has recently reported the detection of an unexpectedly stronger absorption signal in the global 21-cm spectrum around cosmic red-shift $z = 17$, resulting in the significant cooling of the primordial gas. The cooling of the gas  can be expected to occur by considering 
the scattering off the baryons by a small fraction of DM carrying a tiny electromagnetic charge (milli-charged DM). However, it turns out the energy density of milli-charged DM obtained by considering thermal annihilation through massless photons will get overproduced in the range of parameters allowed by 21-cm line signal, if there are no new force carriers etc. In this study, we argue that the milli-charged DM particles might get decoupled during matter domination, in case there are moduli present in the theory. Therefore we estimated the value of relic abundance of milli-charged DM by taking into account an early matter dominated era. We found that for the value of modulus mass around $m_{\Phi} \approx 10^6$~TeV, the milli-charged DM particles could actually yield the desired fraction of DM particles for the values of charge and DM required to explain the 21-cm signal. Thus,  the problem of an overproduction of the milli-charged DM density in the desirable range of mass and charge of DM can be evaded by considering an early moduli/matter dominated era. 
\end{abstract}

\maketitle

\section{Introduction}
Inspite of the rigorous search over decades, the nature of Dark Matter (DM) still remains very puzzling. Although
almost all of our evidence for the existence of DM mainly relies on its gravitational interactions with an ordinary baryonic matter, we expect that it should have some level of non-gravitational interaction with standard-model (SM) baryons. 
In fact, there exist a few puzzles at small scale of the universe which require DM to have non-zero gravitational interactions \cite{Mateo:1998wg,Spergel:1999mh}. The attempts to probe the exact nature of DM are being carried out intensively in a wide range of direct and indirect detection experiments, accelerator-based searches, and some of the cosmological observations. However, unfortunately, none of these experiments has been able to find a conclusive evidence for the non-gravitational interactions of DM till now.
 
An interesting novel probe to search for non-gravitational interactions of DM with ordinary baryonic gas relies on studying the red-
shifted 21 cm signal of neutral hydrogen in the cosmic dawn. The first detection of the 21-cm signal from cosmic dawn has been reported recently by an Experiment to Detect the Global Epoch of reionization Signature (EDGES)~\cite{Bowman:2018yin}. After removing the instrumental noise and the foregrounds, the extracted  signal features a broad absorption dip with brightness temperature  $T_{21} = -500^{+200}_{-500}$ mK, centred at redshift $z = 17.2$. In the standard cosmological model, the maximum possible brightness temperature expected at $z=17$ is around $T_{21} \sim-209$~mK. This value corresponds to a baryon gas temperature $T_b \sim 7$~K. However, the observed $T_{21} = -500$ mK  requires the primordial gas to be much colder, $T_b \sim 4$~K, which is difficult to explain by astrophysical models alone~\cite{Bowman:2018yin}.  It has been argued in Ref.~\cite{Barkana:2018lgd,Fialkov:2018xre} that the observed discrepancy can be accounted by considering the scattering off the baryons by light DM, thus draining excess energy from the primordial gas. The significant cooling at a particular red-shift can follow from the elastic scattering between DM and baryons with a Rutherford cross section with $1/v^4$ dependence, where v is the relative velocity between DM and baryons ~\cite{Tashiro:2014tsa,Munoz:2015bca}. In the context of particle physics models, the $v^{-4}$ cross section corresponds to DM interacting with baryons either through a massless photon or a light mediator with mass smaller than the momentum transfer. However, the tight constraints from $5^{\rm th}$ force experiments and limits from stellar cooling completely rule out the possibility of light mediators \cite{Barkana:2018qrx}. Several studies have explored the milli-charged DM interpretation of the EDGES signal by considering its annihilation through the SM photons \cite{Munoz:2018pzp,Berlin:2018sjs,Kovetz:2018zan}, and an additional massive gauge boson \cite{Berlin:2018sjs,Barkana:2018qrx,Klop:2018ltd,Kovetz:2018zes}.  For the case of milli-charged DM interacting through SM photons, it has been found in Refs.~\cite{Munoz:2018pzp,Berlin:2018sjs} that the viable parameter space of 21-cm signal can be obtained  if only a small fraction of the DM ($f_{\rm DM} \approx 0.02$) is charged. However, the relic density of the milli-charged DM obtained by standard thermal annihilations gets overproduced in the desired region of parameter space.

In this work, we  explore the possibility that an overproduced relic density of the milli-charged DM might be depleted by taking into account an early matter/moduli dominated era. Moduli are the massless scalar fields, which appear generically in the consistent string compactifications~\cite{Coughlan:1983ci,Banks:1993en,deCarlos:1993wie,Dine:2000ds,Acharya:2008bk,Kane:2015jia}. Their presence generically give rise to a period of matter dominated era in the early universe. This allows for several new possibilities of the production of DM in the early universe. We argue that for the desirable range of parameters of DM, it is likely for the DM to get decoupled during matter dominated era. Since the expansion rates are different for both matter and radiation dominated era, this would certainly yield different values of the final relic abundance of milli-charged DM. Thus,  we calculate the relic abundance of milli-charged DM in the modulus dominated era, and show that both the desired fraction of DM relic density and the absorption signal of 21-cm can be explained for the similar values of mass and charge of DM. Quite interestingly, this 
sets a bound on the mass of modulus $m_{\Phi} \approx 10^7~{\rm GeV}$,  which exactly matches with the value of modulus mass given in one of the promising string compactification scenarios, dubbed as Large Volume Scenario (LVS) \cite{Balasubramanian:2005zx,Blumenhagen:2009gk,Cicoli:2012aq,Aparicio:2014wxa} in order to obtain the observed inflationary predictions and ${\cal O}(\rm TeV)$ scale supersymmetry breaking.
  
 The paper is structured as follows: In $\$$\ref{sec:21cm}, we discuss the constraints obtained on the mass and charge of milli-charged DM from the measurements of 21-cm line signal reported by EDGES collaboration, along with several experimental and astrophysical constraints. In $\$$\ref{sec:DM and mod dom}, we  briefly discuss the dynamics of the moduli present in the early universe, and the possibility of milli-charge DM getting decoupled in an  early modulus/matter dominated era. In $\$$\ref{subsec:relic}, we present the detailed derivation of calculating relic abundance of DM in matter dominated era. In $\$$\ref{subsec:dilution}, we give various expressions useful to calculate the amount of dilution in the energy density of DM particles after the decay of modulus.   In $\$$\ref{sec:depletion}, we apply our results of the relic abundance obtained during matter domination to a case with milli-charged DM.  In $\$$\ref{subsec:constraints}, we give constraints on the mass of the modulus. In $\$$\ref{subsec:ps}, we show different regions in the parameter space of mass and charge of milli-charged DM allowed from the requirement of explaining the EDGES 21-cm line signal, and obtaining  a particular fraction of milli-charged DM relic density in both matter and radiation dominated era. The results show that unlike radiation dominated era,  it is possible to obtain the desirable fraction of the milli-charged DM relic density in the range of parameters allowed by 21-cm line signal in the matter/modulus dominated era. In  $\$$\ref{sec:summary}, we summarize our results.

\section{Milli-charged DM and 21-cm Line Signal}
\label{sec:21cm}
It was discussed in Ref.~\cite{Munoz:2018pzp} that the required 21-cm absorption features can be explained if the milli-charged DM forms only a small fraction of the entire DM of the universe.   After numerically solving the Boltzmann equation for the temperature of baryon gas (by taking into effect the interaction of DM with baryons), they have found out that the baryon temperature $T_b \approx 4$K can be obtained if the charge of DM satisfies the following constraint:
\beq
\epsilon \approx 6 \times 10^{-7} \left(\frac{m_X}{\rm MeV}\right) \left(\frac{f_{\rm DM}}{10^{-2}}\right)^{-3/4},
\eeq
where $f_{\rm DM}$ is the fraction of the DM possessed by milli-charged particles. In addition to the constraint on the charge of DM, in order for the DM  to cool the baryonic gas efficiently, the equipartition theorem requires that the mass of dark matter particles should be $m_X < 6.2~{\rm GeV} \times f_{\rm DM}$.
 
Now, the parameters of milli-charged DM are also subjected to the constraints from various different experiments. Following Refs.~\cite{Munoz:2018pzp,Berlin:2018sjs}, we take into account the constraints from observations of supernovae SN1987A~\cite{Chang:2018rso}, SLAC milli-charge experiment~\cite{Prinz:1998ua}, bounds on the abundances of light elements  produced during Big Bang Nucleosynthesis (BBN)~\cite{Boehm:2013jpa}, constraints
from measurements of the CMB on the DM annihilation in the epoch of recombination~\cite{Ade:2015xua},  and on DM scattering with baryons~\cite{Xu:2018efh}.  The constraints on dark matter scattering with baryons during kinetic decoupling (KD) in Ref.~\cite{Xu:2018efh}, and on DM annihilation in the epoch of recombination in Ref.~\cite{Ade:2015xua} are presented for $f_{\rm DM}=1$. Based on the analytical expressions given in Ref.~\cite{Xu:2018efh,Ade:2015xua}, we have rescaled the constraints on DM annihilation constraints by a factor of $f^{2}_{\rm DM}$ and on DM scattering cross-section by a factor of $\sqrt{f_{\rm DM}}$. As discussed in \cite{Berlin:2018sjs}, the constraints from the DM scattering with baryons would not apply for the case of $f_{\rm DM}\lesssim 0.01$ because the energy density of the milli-charged  DM is less than the difference between the upper limit (at 95$\%$~
CL) imposed on the baryon density from CMB and the lower limit (at 95$\%$~CL) imposed on the baryonic density from BBN.

We have shown all of the aforementioned constraints in the form of different shaded regions in Fig.~1. In this figure, we have also shown the region in the parameter space required to explain the $21$-cm absorption feature as reported by EDGES collaboration, along with the parameter space allowed from the requirement of obtaining the desired fraction of milli-charged DM abundance in both matter and radiation dominated era.  Our results in Fig.~1 would clearly point out that that the issue related to the overproduction of DM density \cite{Berlin:2018sjs} in the desirable range of mass and charge of DM  can be avoided if the DM is produced in an early matter dominated era. Before we discuss our results in $\$$~\ref{sec:depletion}, we will first derive the expression to calculate the relic abundance of DM in matter/modulus dominated era. The readers not interested in going through steps of the derivation of DM relic abundance can  directly  use the final expression of DM relic abundance in Eq.~(\ref{eq:omegaDM}), and then move to the final results in $\$$~\ref{sec:depletion}.

\section{Moduli dominated Cosmology}
\label{sec:DM and mod dom}
The moduli are the massless scalar fields mostly appearing in string compactifications. Because of their gravitational interactions, they get significantly displaced from the minimum of the potential just after the end of inflation \cite{Coughlan:1983ci,Banks:1993en,deCarlos:1993wie,Dine:2000ds,Acharya:2008bk,Kane:2015jia}. Subsequently, the energy density associated with the  moduli fields begin to redshift like matter. Since the energy density of moduli would scale as $1/a^3$ during the expansion of the universe (in comparison to the energy density of radiation being scaled as $1/a^4$), they will start dominating the universe quickly after the end of inflation, leading to an early period of matter dominated era. Eventually, when the Hubble expansion rate becomes of the order of $\Gamma_{\rm mod }$, it will decay and the universe will again enter in the radiation dominated era. The decay of moduli into relativistic SM particles will increase the entropy of the universe, thus again reheating the universe. The requirement that the decay of the moduli into SM particles shall not disturb the constraints imposed on the abundances of light elements produced by Big Bang Nucleosynthesis (BBN), one has to assume that the moduli decay before $T \le  {\rm MeV}$. For the decay width of moduli given by  $\Gamma_\Phi \sim m^{3}_{\Phi}/{M^2_p}$, its reheating temperature $T_r =\sqrt{{\Gamma_\Phi} M_p}$ will be less than  ${\cal O}(\rm MeV)$ if the moduli mass $m_{\Phi} \geq 10$~TeV \cite{Coughlan:1983ci,Banks:1993en,deCarlos:1993wie,Dine:2000ds,Acharya:2008bk}. 
 
In the standard cosmology, we assume that the universe is dominated by radiation from the point of inflationary reheating until the period of BBN. Therefore,  DM also gets decoupled during radiation domination. Since the inclusion of moduli allow an early period of matter domination, it is not necessary that the DM decoupling will occur during radiation domination.  We consider  a scenario in which early universe is formed of a thermal bath, comprised of the visible sector states, dark matter, and a single moduli field (named as modulus). Depending on the freeze-out temperature of DM ($T_{\rm FO}$), modulus decay temperature ($T_{\Gamma}$), and the temperature at which modulus started dominating ($T_m$), three distinct regimes might occur \cite{Hamdan:2017psw}:
\begin{enumerate}
\item {{\bf Radiation domination:}} If $T_{\Gamma}>T_{FO}$,  DM particles would  get decoupled in the radiation dominated universe. This is the standard radiation domination case. As far as the decoupling of milli-charged DM particles in radiation dominated era is concerned,  it is discussed in refs.\cite{Munoz:2018pzp,Berlin:2018sjs} that for the values of DM parameters required to explain the 21-cm absorption feature, the relic density of DM will always be overproduced. Thus, we will not consider this case.
\item {{\bf Radiation domination with dilution:}} If $T_{FO}>T_{m}, T_{\Gamma}$, DM particles would again decouple during the radiation dominated era. However, since the modulus starts dominating at $T_{m}<T_{FO}$ and then decays at $T_{\Gamma}$, the overall relic density of DM  particles will get diluted because of an increase in the entropy of the universe.  

For the case of milli-charged DM with mass~$\sim{\rm GeV}$, the freeze-out temperature $T_{FO} < {\rm GeV}$.  On the other hand, for $m_{\Phi}\ge 10$ TeV, and $T_m \sim \sqrt{m_{\Phi} M_p}\ge  5 \times 10^6$ TeV (see Eq.~(\ref{eq:Tosc})), we will always get $T_{FO} << T_m$. Thus, milli-charged DM will never fall under this category. 

\item {{\bf Matter domination:}} If $T_{m} <T_{FO} < T_{\Gamma}$, DM particles would decouple during matter domination. Since the expansion rate in the matter domination case scales differently as compared to radiation dominated case, it will lead to a different amount of relic density of milli-charged DM. Since it is likely to have $T_{m} <T_{FO} < T_{\Gamma}$ in the range of the DM parameters allowed by 21-cm line signal, we will consider this possibility in our study.
\end{enumerate}
In the next subsection, we first calculate the relic abundance of milli-charged DM particles by considering an assumption that the same gets decoupled in the matter dominated era. 
 

\subsection{Relic abundance of DM in matter dominated era} 
\label{subsec:relic}
The number density $n_X$ of DM  particles will follow the standard form of the Boltzmann equation given by
\beq
 {\dot n_X} + 3 H n_X = -\langle\sigma v \rangle \left[n^2_X - (n^{\rm eq}_X)^2 \right].
\eeq
where $\langle\sigma v \rangle$ is the thermally averaged cross-section of DM particles. By using $H = {\dot T}/T$, we can re-write the LHS in terms of co-moving number density  given by $Y=n_X/s$. Defining $x=m_{X}/T$ and $\langle\sigma v \rangle = \sigma_0/x^n$, we will have 
\beq
\label{eq:Y}
 H x^{n+1} \frac{dY}{dx} = -\sigma_0 s \left[Y^2_X - (Y^{\rm eq}_X)^2 \right],
\eeq
where $s=\frac{2\pi^2}{45} g_{*} T^3$ is the entropy density. The expansion rate $H$ is generally given by:
\begin{equation}
H^2 = \frac{8\pi}{3 M^2_p}\left(\rho_{\gamma} + \rho_\Phi + \rho_{\rm DM}\right).
\end{equation}
We can parametrise $H=H(T)$ in terms of critical temperature $T_*$  ($T_* > T> M_X)$, and the relative fraction of visible energy density $r=\frac{\rho_{\gamma}+\rho_X}{\rho_{\gamma} + \rho_\Phi + \rho_{\rm DM}}$ as follows:
\beq
\frac{H^2}{H^2_*} = r \left(\frac{a_*}{a}\right)^4 + (1-r) \left(\frac{a_*}{a}\right)^3,
\eeq 
where $H_* = H(T_*)$, $a_* = a(T_*)$. Here, $T_*$ corresponds to the temperature after which matter domination era begins to start.
Since the thermal plasma is formed of radiation, by rewriting $a_*/a$ in terms of $T/T_*$, the Hubble expansion rate will be~\cite{Hamdan:2017psw}:
 \beq
 \label{eq:Hubble}
 {H} = H_{*}  {\left(\frac{g_*(T_*)}{g_*(T)}\right)}^{3/8}  {\left(\frac{T}{T_*}\right)}^{3/2}  \left[(1-r)+ r\left(\frac{T}{T_*}\right)\right]^{1/2}.
\eeq 
Note that for radiation dominated case i.e $r \approx 1$, we reproduce the standard form of $H \propto T^2$, whereas for modulus/matter domination case i.e $r << 1$, it gives $H \propto T^{3/2}$.  By rewriting $Y$ as $\Delta = Y - Y_{\rm eq}$, and the expression of $H$ given in Eq.~(\ref{eq:Hubble}), we can write \beq
\label{eq:Delta}
\Delta^\prime \approx   - Y^\prime_{\rm eq} - \lambda \left(1-r + r \frac{x_*}{x} \right)^{-1/2} x^{-5/2-n} \Delta\left[2 Y_{\rm eq} + \Delta \right],
\eeq
where primed variables correspond to derivatives with respect to $x$, and $\lambda= {2 \pi^2 g_{*S} m^3_X \sigma_0}/{(45 H_* x^{3/2}_*)}$. Since we are interested to obtain $Y$ in the domain $x > x_{FO}$, we will always have $Y > Y_{eq}$. Thus, we can write the late time difference $\Delta \sim Y$, and may also ignore $Y^\prime_{\rm eq}$ in Eq.~(\ref{eq:Delta}). With this, the present time $\Delta_{\infty} = \Delta|_{x > x_{FO}}$ can follow from
\beq
\label{eq:Delta1}
\Delta^\prime_{\infty} \approx   - \lambda \left(1-r + r \frac{x_*}{x} \right)^{-1/2} x^{-5/2-n} \Delta^2_{\infty}.
\eeq
 By integrating the above equation between $x=x_{FO}$ to $x \approx 0$, the freeze-out co-moving number density will be:
 \beq
\label{eq:YFO}
Y_{\rm FO} \approx \left(\lambda   \left(1-r + r \frac{x_*}{x} \right)^{-1/2} x^{-5/2-n} \right)^{-1}.
\eeq

By evaluating the integral for $r<<1$, we would obtain the freeze-out co-moving DM number density in the matter dominated case as  \cite{Hamdan:2017psw}:
 \beq
\label{eq:YFOmatter}
Y^{\rm matter}_{\rm FO} \approx \left(n+\frac{3}{2}\right)\frac{x^{n+\frac{3}{2}}_{FO}}{\lambda} = 3 \frac{\sqrt{5}}{{\pi}}\frac{\sqrt{g_*}}{g_{*S}}\frac{\left(n+\frac{3}{2}\right)x^{n+\frac{3}{2}}_{FO}}{M_p m_X \sigma_0 \sqrt{x_*}}.
\eeq
Comparing this with the standard expression of $Y_{FO}$ in radiation dominated case, we can figure out that (i) $Y^{\rm matter}_{\rm FO}$ is proportional to $x^{n+\frac{3}{2}}_{FO}$, instead of $x^{n+1}_{FO}$ in radiation dominated case, (ii) this depends on an additional parameter $x_*$, which further depends on the details of dynamics of modulus in the early universe.
\\
\paragraph*{Freeze-out temperature:} For the number density of non-relativistic DM particles given by $n_X = \frac{g_X}{(2\pi)^{3/2}}m^3_X x^{-3/2} e^{-x}$,  the freeze-out temperature can found by equating $H (T_{FO}) = n_{X} \langle \sigma v \rangle$. By using the expression of H given in Eq.~(\ref{eq:Hubble}), it will turn out to be  \cite{Hamdan:2017psw}:
\beq
x_{\rm FO} ={\rm \ln}\left[\frac{g_X m^{3}_X \sigma_0  }{(2\pi)^{3/2} x^{3/2}_{*}H_{*}}  \left(\frac{x^{-1+2n}_{FO}}{(1-r)x^{4n-1}_{FO} + x_* r }\right)^{1/2} \right],
\eeq
where $H_{*}=H_{*}(T_*)$ and $x_* =m/T_*$.  By taking $r<<1$ and $H_* = {8 \pi^3 g_* T^4_*}/{M^2_p}$, for $n=0$, it gives the following value of $x_{FO}$ in matter domination case. 
\beq
x^{\rm MD}_{\rm FO} \approx \ln\left[\frac{3}{4 \pi^3} \sqrt{\frac{5}{3}} \frac{g_X}{g_*}\frac{m^{3/2}_X M_p\sigma_0}{T_*}\right].
\eeq
This expression will be used while calculating the value of final relic abundance of DM.
\subsection{Dilution of relic density}
\label{subsec:dilution}
 As discussed in \ref{sec:DM and mod dom}, the decay of modulus into relativistic particles might increase the entropy of the universe at late time, thus again reheating the universe. Therefore, after the decay of modulus, the co-moving number density of DM particles calculated during matter dominated era might get diluted. The final freeze-out co-moving number density will be:
\beq
Y^{\rm matter}_{FO}|_{\rm after} = Y^{\rm matter}_{FO}|_{\rm before} \left(\frac{s_{\rm after}}{s_{\rm before}}\right) = \frac{Y^{\rm matter}_{FO}|_{\rm before} }{\xi},
\eeq
where $\xi=\frac{s_{\rm after}}{s_{\rm before}}$ is the dilution factor. Now we summarize the expressions of decay and reheating temperature in order to estimate the increase in the entropy \cite{Acharya:2008bk}. The decay temperature can be found as follows:
\beq
3 H^2_{decay} = \frac{m_{\Phi} Y_{\Phi} s_{\rm decay}}{M^2_p} = \frac{m_{\Phi} Y_{\Phi}}{M^2_p}\frac{2 \pi^2}{45} g_{*S}(T_d) T^3_d,
\eeq
where $Y_{\Phi} = n_{\Phi}/s$ is the co-moving number density of modulus. Its value will be given as: 
\beq
\label{eq:Yphi}
Y_{\Phi} = \frac{1}{2} m_{\Phi} f^{2}_{\Phi} s^{-1} (T_{\rm osc}),
\eeq
where $f_{\Phi}$ is the oscillation length of modulus, which will be typically of an ${\cal O}(M_p)$, and $T_{\rm osc}$ is the temperature at which the modulus starts oscillating. The value of $T_{\rm osc}$ will follow from:
\bea
\label{eq:Tosc}
&& 3 H^2 = 3  \left(\frac{1}{2} m_{\Phi}\right)^2 M^2_p = M^{-2}_p \left(\frac{\pi^2}{30}\right) g_{*} (T^{\Phi}_{\rm osc})^4 \nonumber\\
&& \rightarrow T^{\Phi}_{\rm osc} = \left(\frac{90}{4 \pi^2} \right)^{1/4} \left(m^2_{\Phi} M^2_p\right)^{1/4}.
\eea
For $H_d =\Gamma_{\rm mod}$, the decay temperature $T_d$ will be given as:
\beq
\label{eq:Td}
T_d = \left(\frac{30}{\pi^3}\right)^{1/3} \left(\frac{\Gamma^2_{\rm mod} M^2_p}{m_\Phi Y_\phi g_{*S}(T_d)}\right)^{1/3}.
\eeq
The reheating temperature after the decay of modulus can be calculated from:
\bea
\label{eq:Tr}
&& 3 H^2 = \frac{4 \Gamma^2_\Phi}{3} = \frac{\pi^2}{30} g_*(T_r) \frac{T^4_r}{M^2_p}, \nonumber\\
&& \rightarrow T_r = \left(\frac{40}{\pi^2}\right)^{1/4} g^{-1/4}_*(T_r) \sqrt{\Gamma_\Phi M_p}
\eea
For the decay width of modulus given by $\Gamma_{\Phi}= D_{\Phi} \frac{m^3_{\Phi}}{M^2_p}$, the value of dilution factor will follow from 
\bea
\label{eq:dilution}
&& \xi =  \frac{s_{\rm after}}{s_{\rm before}} \approx \frac{g_{*S} (T_r)}{g_{*S} (T_\Gamma)}\left(\frac{T_r}{T_\Gamma}\right)^3 \nonumber\\
&& = \frac{2}{15} \left( 250 \pi^2 \right)^{1/4} g^{1/4}_{*}\left(\frac{M_p}{D_{\Phi} m_{\Phi}}\right)^{1/2} Y_{\Phi},
\eea
where $Y_{\Phi}$ can be calculated by using Eqs.~(\ref{eq:Yphi}) and (\ref{eq:Tosc}).
\subsubsection{Present DM relic abundance} 
Here we write down the final expression of the present day relic abundance of DM that we can obtain if it gets decoupled in matter dominated era. 
\beq
\Omega^{\rm matter}_{0,DM} = \frac{1}{\xi} \times \frac{ s_0 m_{X} Y_{FO}}{\rho_c}, 
\eeq
where $s_0= \frac{2 \pi^2}{45} g_{*S}(T_0) T^3_0$, with $T_0=2.34 \times 10^{-10}~{\rm MeV}$, and ${\rho_c}=8.05 \times 10^{-35}~{\rm MeV}^4$.  The value of $T_*$ will be same as $T^{\Phi}_{\rm osc}$ calculated in Eq.~(\ref{eq:Tosc}). Using Eq.~(\ref{eq:YFOmatter}) and Eqs.~(\ref{eq:Yphi})-(\ref{eq:dilution}), for $g_* = g_{*S} \approx 100$, the final DM relic abundance will follow from:
\beq
\label{eq:omegamat}
\Omega^{\rm matter}_{0,DM} = \left(\frac{3.5 \times 10^5}{\rm MeV}\right) \frac{1}{\sqrt{m_X}}  \frac{1}{\langle \sigma v \rangle}\left[\frac {m^5_{\Phi}}{M^7_p}\right]^{1/4} .
\eeq

\section{Depleting the relic abundance of milli-charged DM in matter dominated era}
\label{sec:depletion}
In this section, we show how can we obtain the right value of DM relic abundance in the desired region of parameter space of mass and charge of milli-charged DM in the modulus dominated era. As mentioned in \cite{Berlin:2018sjs}, the annihilation cross section for a pair of milli-charged particles is given by:
\beq
\label{eq:sigmav}
\langle \sigma v \rangle = \frac{\pi {\alpha^2} \epsilon^2}{m^2_{X}} k \left(\sqrt{1-\frac{m^2_f}{m^2_X}}\right)\left(1+\frac{m^2_f}{2 m^2_X}\right).
\eeq
where $k=1$ and $v^2/6$ for Dirac fermion and complex scalar respectively, $\epsilon$ corresponds to charge, and $m_f$ corresponds to mass of fermion into which DM particles annihilate. Since we assume that DM is in the form of Dirac fermion, we will take $k=1$. Now, by using Eqs. (\ref{eq:omegamat}) and (\ref{eq:sigmav}), the relic abundance of milli-charged DM particles will follow from:
\bea
\label{eq:omegaDM}
  \Omega^{\rm matter}_{X} & \approx 0.1 & \left[\frac{m_{\Phi}}{10^4~{\rm TeV}}\right]^{5/4} \left[\frac{m_X}{10~\rm MeV}\right]^{3/2}  \nonumber\\
&&   \times \left[\frac{10^{-5}}{\epsilon}\right]^2 \left[\frac{x^{\rm MD}_{\rm FO}}{20}\right]^{3/2}.
\eea
 As discussed in $\$$\ref{sec:21cm}, the milli-charged DM particles are allowed to form only a fraction of the whole DM of the universe. Thus we parametrize the milli-charged DM relic abundance by $\Omega^{\prime \rm matter}_{X}= f_{\rm DM}~\Omega^{\rm matter}_{X}$, where $f_{\rm DM}$ corresponds to the fraction of the total DM. From Eq. (\ref{eq:omegaDM}), we can calculate the range of $\epsilon$ and $m_X$ for which the desired value of relic abundance  ($\Omega^{\prime \rm matter}_{X}$) can be obtained.

   \begin{figure*}[htp]
  \centering
  \hspace*{-1cm} 
 \subfigure[~~Milli-charged DM fraction {\bf  $f_{\rm DM} =1$}]{\includegraphics[width = .48\textwidth,height = .40\textwidth]{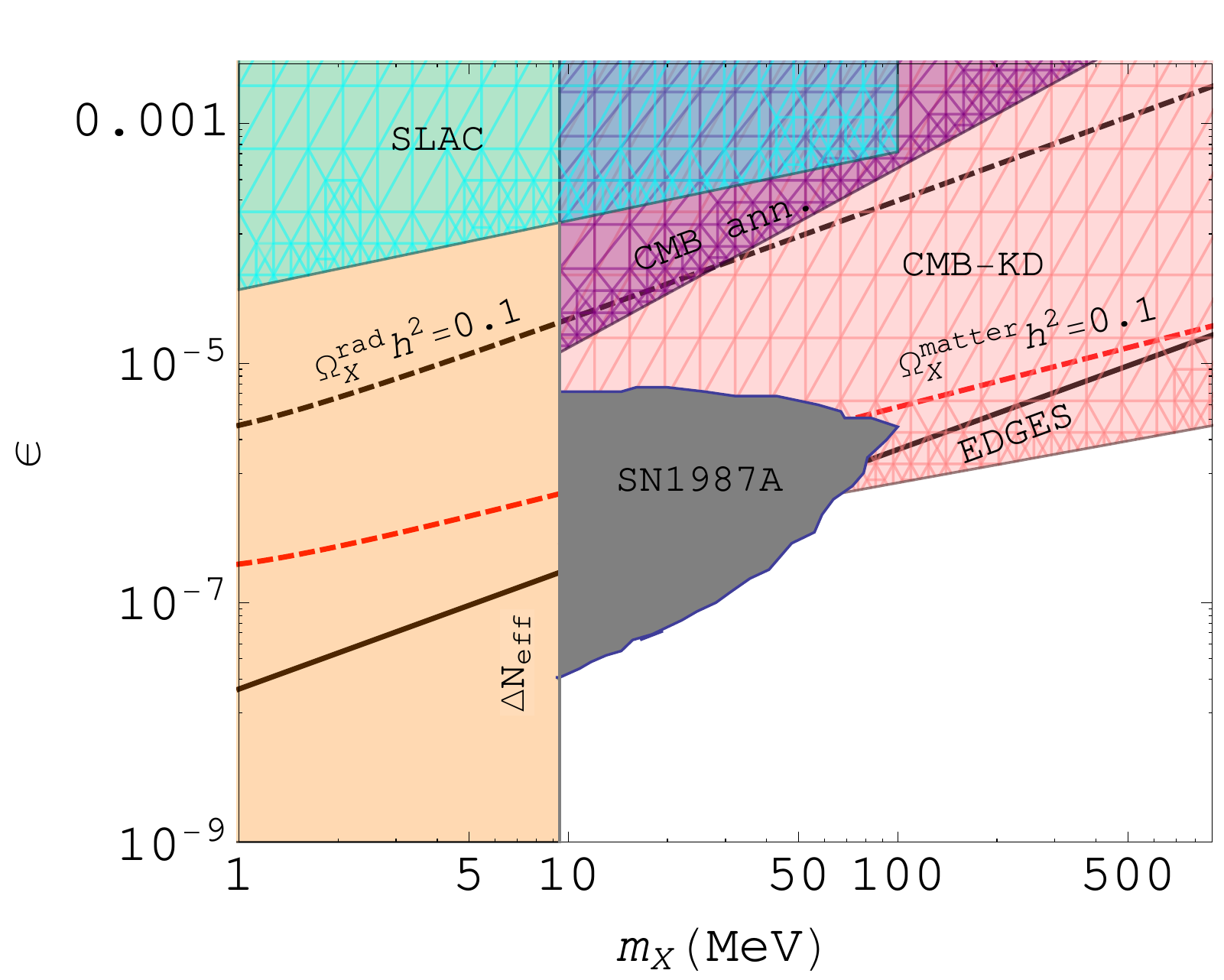}} \quad
  \subfigure[~~Milli-charged DM fraction {\bf   $f_{\rm DM} =0.1$}]{\includegraphics[width = .48\textwidth,height = .40\textwidth]{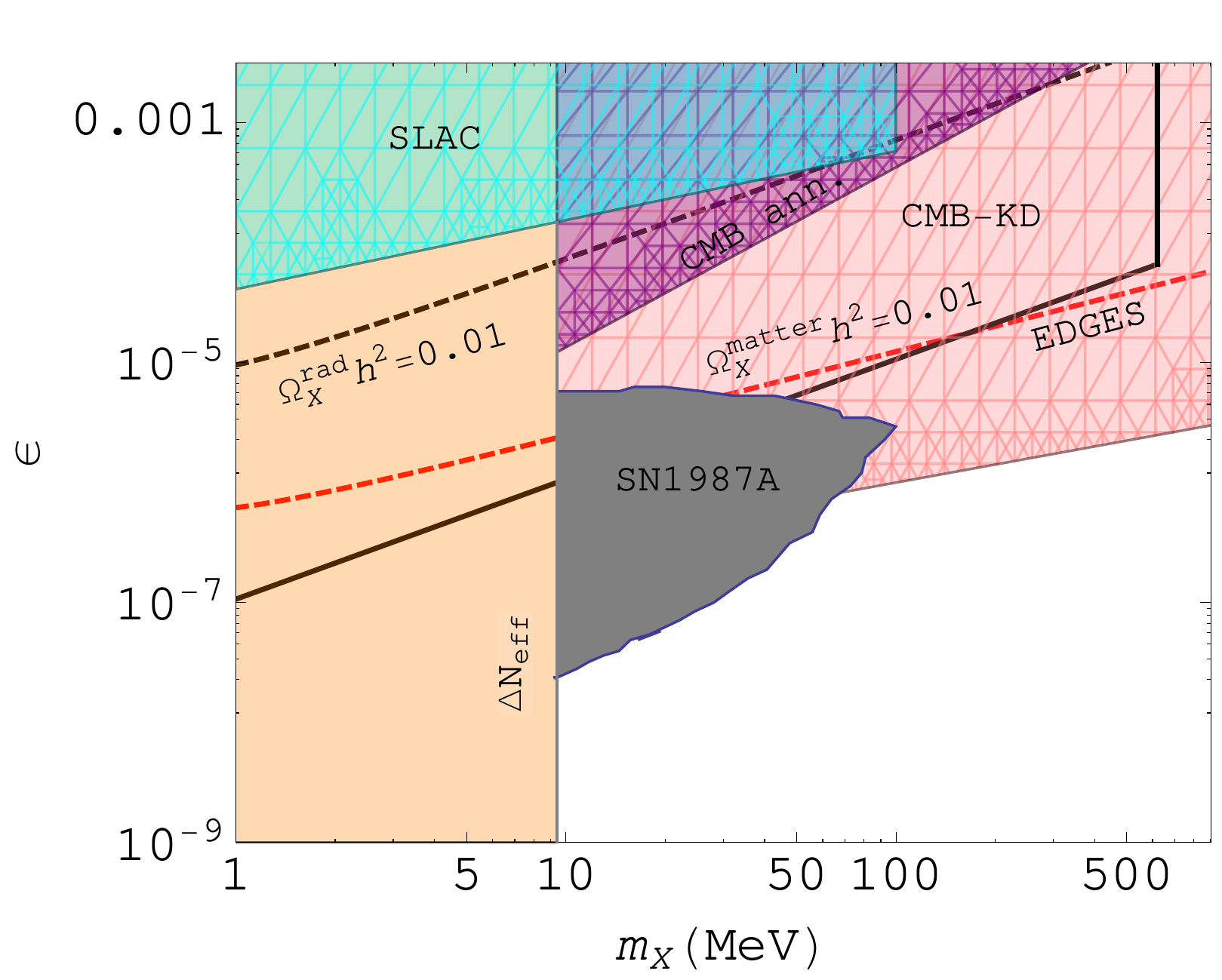}}  
  \hspace*{-1cm} 
  \subfigure[~~Milli-charged DM fraction {\bf   $f_{\rm DM}=0.01$}]{\includegraphics[width = .48\textwidth,height = .40\textwidth]{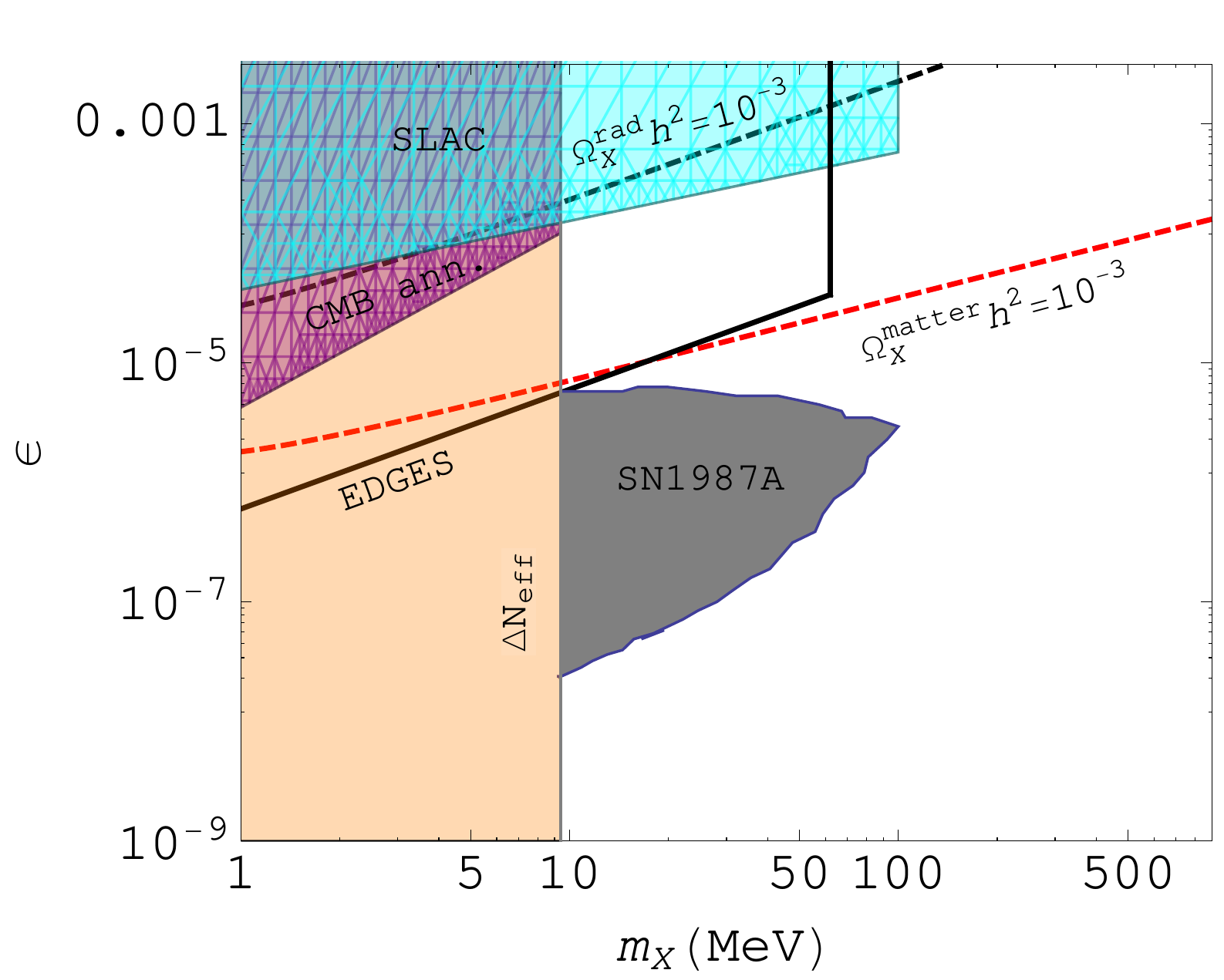}} \quad
   \subfigure[~~Milli-charged DM fraction {\bf   $f_{\rm DM} =0.001$}]{\includegraphics[width = .48\textwidth,height = .40\textwidth]{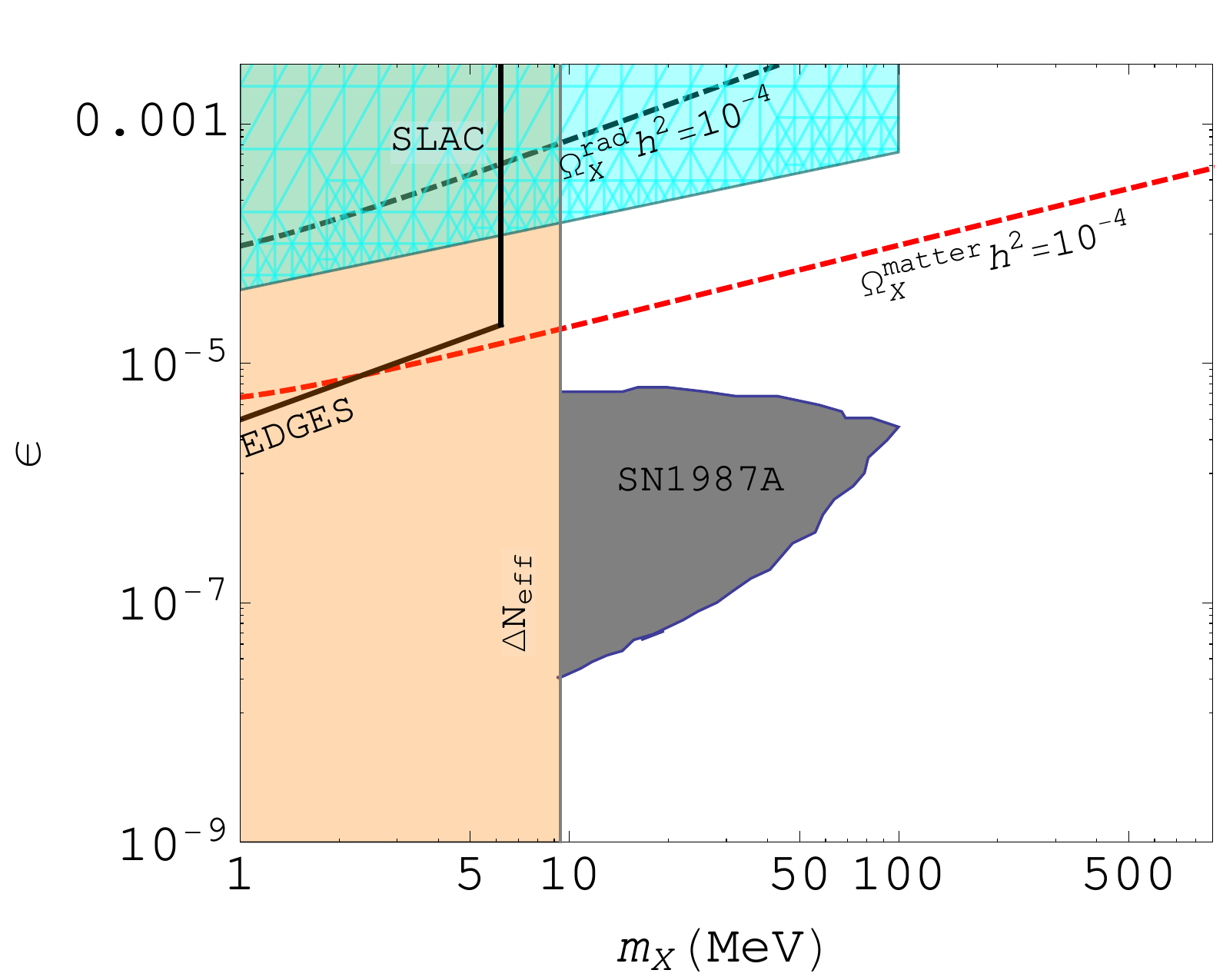}}
  \caption{We show the region in the parameter space for which the desirable value of relic abundance can be obtained by considering an early modulus dominated era. The black solid lines in Figs.~1(a)-1(d) indicate the ranges of DM parameters required to explain the observed 21-cm signal reported by EDGES collaboration. The dashed red line in each figure indicates the parameter space for which the relic abundance of milli-charged DM abundance becomes equal to the quoted fraction $f_{\rm DM}$.  The black dashed line in Figs.~1(a)-1(d) indicates the parameter space required to get a particular fraction of milli-charged DM if the same gets decoupled in radiation dominated era. We also show constraints from Supernova 1987A (dark grey, labeled SN 1987A), the SLAC milli-charge experiment (cyan, labeled SLAC), the production of  light element abundances produced during BBN (light orange, labelled $\Delta N_{\rm eff}$), and the constraints from DM scattering with baryons (pink, labeled CMB-KD)  and DM annihilation (dark purple, labeled CMB ann). The fact that the dashed black lines do not coincide with the solid black lines, while the dashed red line intersects with the solid black line for a particular value of mass and charge of DM in each figure indicates that DM might be produced in an early matter dominated era.}
\end{figure*}
 
\subsection{Constraints on the mass of modulus}
\label{subsec:constraints}
Before we discuss the parameter space of mass and charge of milli-charged DM obtained in modulus dominated era, we will collect the bounds on the modulus mass put by an experimental and other considerations. As discussed in $\$$\ref{sec:DM and mod dom}, 
the lower bound on the mass of modulus ($m_{\Phi} \ge 10$ TeV) is already enforced from the requirement of not spoiling the predictions arising from Big-Bang Nucleosynthesis (BBN).  In order to ensure that the DM gets decoupled in matter dominated era, we have to keep  $T_d < T_{FO}$. Using Eq. (\ref{eq:Td}), for $x_{FO} \approx 10$ and $g_{*S}(T_d) \approx 100$, this requirement boils down to give:
\beq
m_{\Phi} \lesssim \left[10^{4} \left(\frac{m_X}{\rm MeV}\right)^{6/11}\right] {\rm TeV}.
\eeq
Since we are interested in the mass range: $1~{\rm MeV} < m_X < 1~{\rm GeV}$, the mass of modulus should always be less than $10^4$ TeV. Thus, we have got the following bound on the mass of modulus:
\beq
10 < \left(\frac{m_{\Phi}}{\rm TeV}\right) < 10^4.
\eeq
\subsection{Allowed Parameter Space}
\label{subsec:ps}
 In Figure 1, we show the parameter space of $\epsilon$ and $m_X$ in which the $\Omega^{\prime \rm matter}_{X}$ forms a particular fraction $f$ of the whole DM of the universe. 
 We have fixed value of modulus mass $m_{\Phi} = 5\times 10^4$ TeV (ensuring that DM gets decoupled in matter dominated era) in all the cases shown in Fig.~1. The black solid line in Figs.~1(a)-1(d) indicates the range of DM parameters required to explain the observed 21-cm signal reported by EDGES collaboration. The dashed red line  indicates the parameter space in which the milli-charged DM abundance obtained in matter dominated era corresponds to the quoted fraction $f_{\rm DM}$ of the whole DM of the universe.  We also compare our results with the parameter space allowed in standard radiation dominated era.  The black dashed line in Figs.~1(a)-1(d) indicates the parameter space required to get a particular fraction of milli-charged DM if the same gets decoupled in radiation dominated era. It can be easily figured out that the  the dashed black line lies quite above the solid black line in the allowed range of parameter space, while the dashed red line coincides with the solid black line for a particular point in the parameter space for all the cases presented in Fig.~1. From this,  we can realize that the desirable milli-charged DM abundance can produced in the allowed range of DM parameters required to explain the 21-cm line signal
 if the DM get decoupled in matter/modulus dominated era. In other words, we can deplete the overproduced relic abundance of milli-charged DM (obtained by considering radiation) if we consider the possibility of DM getting decoupled in matter dominated era.

Now, the parameters of milli-charged DM are also  subjected to various experimental and astrophysical constraints.  All the constraints are summarised  in $\$$\ref{sec:21cm}. The shaded regions in Fig.~1 represent the regions ruled out by various experimental constraints. By taking these into account, we can realize that the region allowed to explain the 21-cm absorption feature, along with the desirable relic abundance, is ruled out completely for the case of $f_{\rm DM} = 1,0.1$ and $0.001$.  However, for $f_{\rm DM} = 0.01$, there is a narrow range of parameter space in which one can explain the EDGES 21-cm absorption feature as well as the desirable fraction of the milli-charged DM.  

Thus,  we are left with a very narrow range of parameter space i.e  $f_{\rm DM} = 0.01$, $10~{\rm MeV} \lesssim m_{X} \lesssim 70~{\rm MeV}$, and $10^{-6} \lesssim \epsilon  \lesssim 10^{-5}$. In Fig.~2, we show our results of DM relic abundance in this narrow range of parameter space. The black solid line shows the parameter space of fraction of the milli-charged DM $f_{\rm DM}=0.01$ required to explain the 21-cm line signal.  The dashed red, dashed pink and dashed brown lines indicate the value of $\Omega^{\rm MD}_X \sim 0.001$ for $m_{\phi} = 4 \times 10^3$~TeV, $6 \times 10^3$~TeV and $10^4$~TeV  respectively.  It appears that by slightly changing the mass of the modulus, it is possible to have the desirable fraction of milli-charged DM abundance in the entire allowed (although narrow) region of the  21-cm line signal. Thus, the region allowed by 21-cm line signal sets a constraint on the modulus mass $m_{\Phi} \approx (4-10)\times 10^3$ TeV.

\begin{figure} 
 \includegraphics[width = .45\textwidth,height = .4\textwidth]{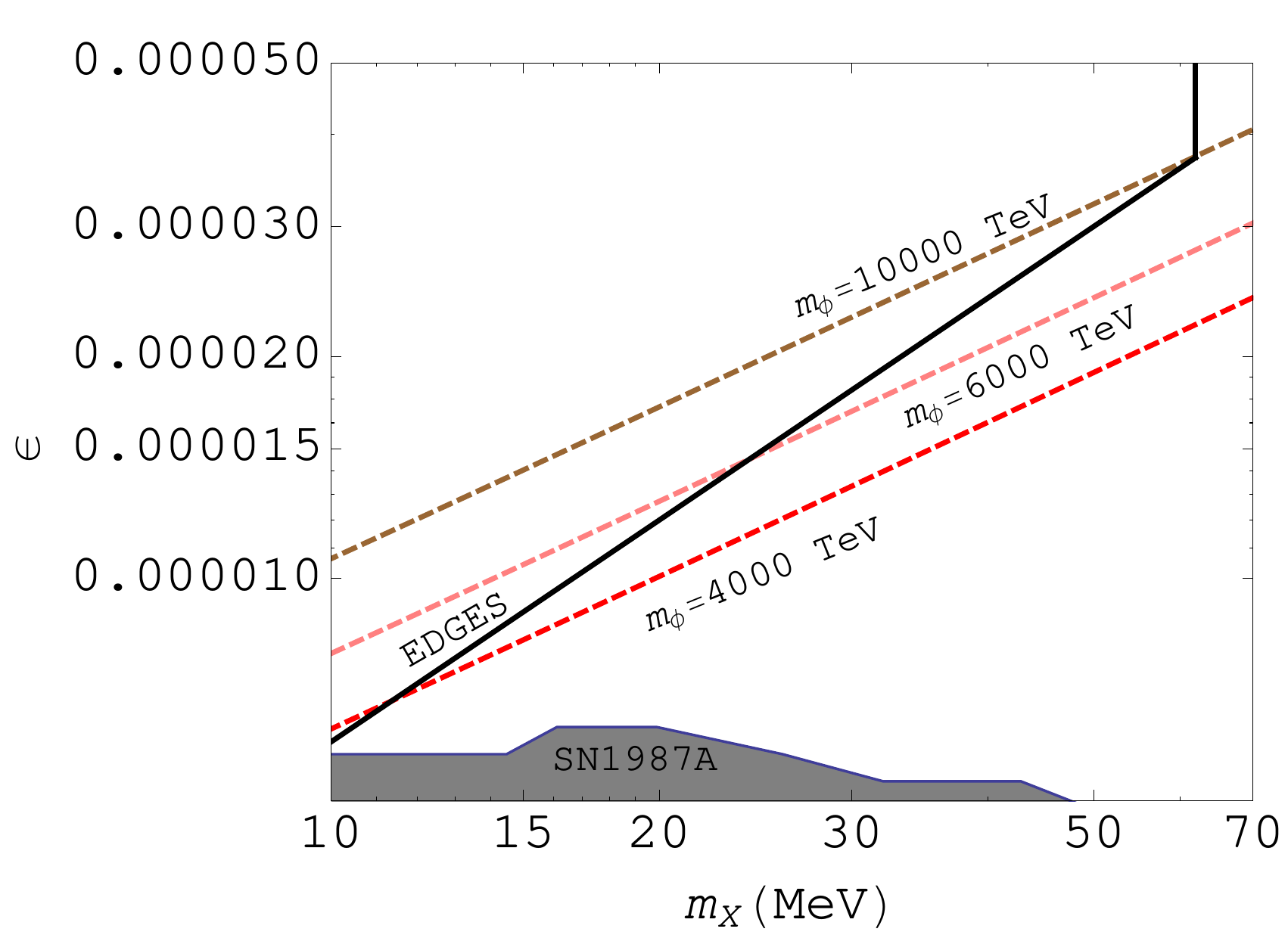}
   \caption{Here we show our results of DM relic abundance in the narrow range allowed by EDGES 21-cm absorption feature. The black solid line shows the parameter space of fraction of the milli-charged DM $f_{\rm DM}=0.01$ required to explain the 21-cm line signal. The dashed red, dashed pink and dashed brown lines indicate the value of $\Omega^{\rm MD}_X \sim 0.001$ (i.e $f_{\rm DM}=0.01$) for $m_{\phi} = 4 \times 10^3$~TeV, $6 \times 10^3$~TeV and $10^4$~TeV  respectively. The results show that by slightly changing the mass of the modulus, it is possible to have the desirable fraction of milli-charged DM abundance in the entire allowed (although narrow) region of the  21-cm line signal. Thus, the region allowed by 21-cm line signal sets a constraint on the modulus mass $m_{\Phi} \approx (4-10)\times 10^3$ TeV. }
\end{figure}

With this, we show that the overproduction of the DM  in requisite range of mass and charge of DM mentioned in Ref.~\cite{Berlin:2018sjs} can be evaded if we take into account the presence of a modulus dominated era in the early universe.
\\


\section{Concluding Remarks}
\label{sec:summary}

Previous studies based on the milli-charged DM explanation of the EDGES 21-cm absorption signal argue that that the energy density of milli-charged DM ($f_{\rm DM} \Omega_X$) obtained by considering thermal annihilation through massless photons will over-close the universe for  the range of parameters allowed by 21-cm line signal 
\cite{Munoz:2018pzp,Berlin:2018sjs}. The overproduced relic density can be depleted by considering extra force carriers, such as $U(1)_{L_{\mu}-L_{\tau}}$ gauge group. However, the yet unconstrained tight parameter space allowed in this case is expected to be within the reach of future measurements \cite{Berlin:2018sjs,Klop:2018ltd}.  Thus one needs a mechanism to deplete the overproduced relic abundance of milli-charged DM. In this article, we have investigated the possibility of depleting the energy density of milli-charged DM particles by taking into account the existence of an early moduli dominated era. Moduli are a ubiquitous prediction resulting from the consistent string compactifications \cite{Kane:2015jia}. Since the successful UV completion of any inflationary model requires its embedding in string theory, we are compelled to consider the presence of moduli in the early universe. The dynamics of moduli tend to change the standard thermal history of the universe, leading to an early matter dominated era. Interestingly, they also allow for a richer range of possibilities for the production of DM in the early universe. Based on the values of mass and charged allowed by 21-cm line signal, we realize that that the milli-charged DM might get decoupled in an early matter dominated era in our case. Thus, we have calculated the relic abundance of milli-charged DM particles for this case. The energy density of DM particles obtained in this case is expected to be less than the one obtained in standard radiation dominated case because  (i) the expansion rate of the universe is different for matter and radiation dominated era, (ii) a significant fraction of the energy density will get diluted once the modulus decays before the onset of BBN. We found that for the value of modulus mass  around $m_{\Phi} \approx (4-10)\times 10^3$ TeV, the milli-charged DM particles could actually yield the desired fraction of DM particles in the region allowed by 21-cm line signal reported by EDGES collaboration. Thus, the overproduction of relic density can be circumvented in case of an early modulus dominated era.  

We have discussed the case of the milli-charged DM interacting with baryons through SM photons. However, it can be extended easily for a case with a light massive mediator. Furthermore, quite interestingly, the exactly same value of $m_{\Phi}$ is obtained  in one of the promising string compactification scenario, dubbed as Large Volume Scenario (LVS) \cite{Balasubramanian:2005zx,Blumenhagen:2009gk,Cicoli:2012aq,Aparicio:2014wxa} in order to obtain the observed bounds on the inflationary predictions and ${\cal O}(\rm TeV)$ scale supersymmetry. Thus,  the results from 21-cm experiments might provide a new window into the dark sector, which might give hints of non-gravitational interaction of DM and ${\cal O}(\rm TeV)$ scale supersymmetry in future .

\section{Acknowledgments}
 The work of MD is supported by Department of Science and Technology, Government of India under the Grant Agreement number: IFA18-PH215 (INSPIRE Faculty Award). MD would like to acknowledge Vikram Rentala (IIT Bombay, India) for several useful discussions on the EDGES 21-cm signal. MD would also like to thank Gaurav Goswami (Ahmedabad University, India) for some useful clarifications.

\end{document}